\documentclass[3p,times,procedia]{elsarticle}
\usepackage{nupha_ecrc}
\usepackage{xspace}
\usepackage{fancybox}
\usepackage{wallpaper} 
\usepackage{multimedia}
\usepackage{wrapfig} 
\usepackage{amsmath}
\usepackage{amsthm}
\usepackage{amsfonts}
\usepackage{subfigure}
\usepackage[latin1]{inputenc}
\usepackage{wasysym}
\usepackage{color}
\usepackage{xcolor}
\usepackage{xspace}
 \usepackage{tikz}
\usepackage{mathtools}
\usepackage{graphicx}

\volume{00}

\firstpage{1}

\journalname{Nuclear Physics A}

\runauth{}

\jid{nupha}

\jnltitlelogo{Nuclear Physics A}

\usepackage{amssymb}

\usepackage[figuresright]{rotating}

\newcommand{\kT}{\ensuremath{k_{\text{T}}}\xspace}

\newcommand{\pT}{\ensuremath{p_{\text{T}}}\xspace}

\newcommand{\dEdx}{\ensuremath{\text{d}E/\text{d}x}\xspace}

\newcommand{\snn}{\ensuremath{\sqrt{s_{_{\text{NN}}}}}\xspace}

\newcommand{\pPb}{\ensuremath{p\text{+Pb}}\xspace}
\newcommand{\pp}{\ensuremath{pp}\xspace}

\newcommand{\avgNpart}{\ensuremath{\langle N_{\text{part}} \rangle}\xspace}

\newcommand{\rSide}{\ensuremath{R_{\text{side}}}\xspace}
\newcommand{\rLong}{\ensuremath{R_{\text{long}}}\xspace}
\newcommand{\rOut}{\ensuremath{R_{\text{out}}}\xspace}

\newcommand{\qSide}{\ensuremath{q_{\text{side}}}\xspace}
\newcommand{\qLong}{\ensuremath{q_{\text{long}}}\xspace}
\newcommand{\qOut}{\ensuremath{q_{\text{out}}}\xspace}
\newcommand{\qInv}{\ensuremath{q_{\text{inv}}}\xspace}
\newcommand{\rInv}{\ensuremath{R_{\text{inv}}}\xspace}

\newcommand{\der}{\ensuremath{\text{d}}\xspace}
\newcommand{\AvgMult}{\ensuremath{\langle \text{d}N_{\text{ch}} / \text{d}\eta \rangle}\xspace}
\newcommand{\SqAvgMult}{\ensuremath{\langle \text{d}N_{\text{ch}} / \text{d}\eta \rangle ^{1/3}}\xspace}

\begin{document}

\begin{frontmatter}



\dochead{}

\title{Femtoscopic measurements in \pPb collisions at $\sqrt{s_{_{\text{NN}}}}= 5.02$~TeV with ATLAS at the LHC}


\author{Markus K. K\"{o}hler \\ (on behalf of the ATLAS\fnref{col1} Collaboration)}
\fntext[col1] {A list of members of the ATLAS Collaboration and acknowledgements can be found at the end of this issue.}
\runauth{M.~K.~K\"{o}hler et al.}

\address{Weizmann Institute of Science, Israel}

\begin{abstract}
Recent measurements in two-particle correlations in $\pPb$ collisions suggest collective behavior reminiscent 
of that observed in Pb+Pb. Femtoscopic measurements may provide useful insight on this behavior because they image the 
spatio-temporal size of the particle emitting region. This proceeding presents identical-pion Hanbury Brown and Twiss 
(HBT) measurements from ATLAS using one- and three-dimensional correlation functions. Pions are identified using $\dEdx$ 
measured in the pixel detector. Correlation functions and the resulting HBT radii are shown as a function of pair 
momentum ($\kT$) and collision centrality.
\end{abstract}


\end{frontmatter}


\section{Introduction}
\label{sec:introduction}
Long range two-particle correlation functions in relative pseudorapidity and relative azimuthal angle in \pp and \pPb 
collisions measured at the LHC are consistent with a possible collective behaviour~\cite{CMS_ridge_pp,CMS_ridge_pPb, 
ATLAS_ridge_pp}, reminiscent to that observed in Pb+Pb collisions, see e.g.~\cite{ALICE_ridge_PbPb}. \\
Femtoscopic measurements in \pp and \pPb collisions~\cite{ATLAS_femto_pp,ALICE_femto_pPb} allow the investigation of 
the time evolution of small colliding systems. This proceeding reports the first measurement of the centrality and 
momentum dependence of same- and opposite-sign pion pair correlations in \pPb collisions at $\snn=5.02$~TeV measured by 
the ATLAS experiment.

\section{Data analysis}
\label{sec:data_analysis}
This analysis uses a data sample with an integrated luminosity of $28.1$~nb$^{-1}$ of \pPb collisions, measured in 2013 
by the ATLAS detector~\cite{ATLAS_detector}. The Pb beam had an energy of $1.57$~TeV per nucleon and the opposing $p$ 
beam had an energy $4$~TeV, resulting in a center of mass energy of $\snn = 5.02$~TeV. The centrality is determined 
using the total transverse energy measured in the forward calorimeter on the Pb-going side, see 
e.g. Ref.~\cite{ATLAS_cent_pPb} for more information. \\
Reconstructed tracks are required to have a transverse momentum $\pT > 0.1$~GeV and to be within the 
pseudorapidity\footnote{ATLAS uses a right-handed coordinate system with its origin at the nominal interaction point 
(IP) in the centre of the detector and the $z$-axis along the beam pipe. The $x$-axis points from the IP to the centre 
of the LHC ring, and the $y$-axis points upward. Cylindrical coordinates $(r,\phi)$ are used in the transverse plane, 
$\phi$ being the azimuthal angle around the $z$-axis. The pseudorapidity is defined in terms of the polar angle $\theta$ 
as $\eta=-\ln\tan(\theta/2)$.} range $|\eta| < 2.5$. Pions are identified using the $\dEdx$ information from the silicon 
pixel detector. Pion pairs are required to have $|\Delta \phi | < \pi/2$ and to be within the pair rapidity region of 
$|\eta_k| < 1.5$. Opposite-sign pairs that have an invariant mass close the masses of $\rho^0$, $K_S^0$ and $\phi$ are 
rejected. \\
The two-particle correlation function is defined as $ C(q) = A(q)/B(q)$, where $q$ is the relative momentum $q = 
p^a - p^b$, $\left. A(q) = \der N / \der q \right |_{\text{same}}$ is the same-event distribution and $\left. B(q) 
= \der N / \der q \right |_{\text{mixed}}  $  is the mixed-event distribution in the same event class.   \\
In three dimensions, a longitudinally co-moving frame is chosen, such that $p_z^a = -p_z^b$. The 
coordinates are defined in the Bertsch-Pratt convention~\cite{Bertsch89,Pratt86}, where $\qOut$ shows in the direction 
of the pair momentum, $\qLong$ shows in the direction of the beam and $\qSide$ is perpendicular to the $\qOut$ and 
$\qLong$. A graphical visualization of Bertsch-Pratt coordinates is shown e.g. in Ref.~\cite{Lisa05}. Using the 
Bowler-Sinyukow~\cite{Bowler91,Sinyukow98} parametrization, the correlation function can written as
\begin{figure}[t]
\begin{minipage}{14pc}
\includegraphics[scale = 0.43]{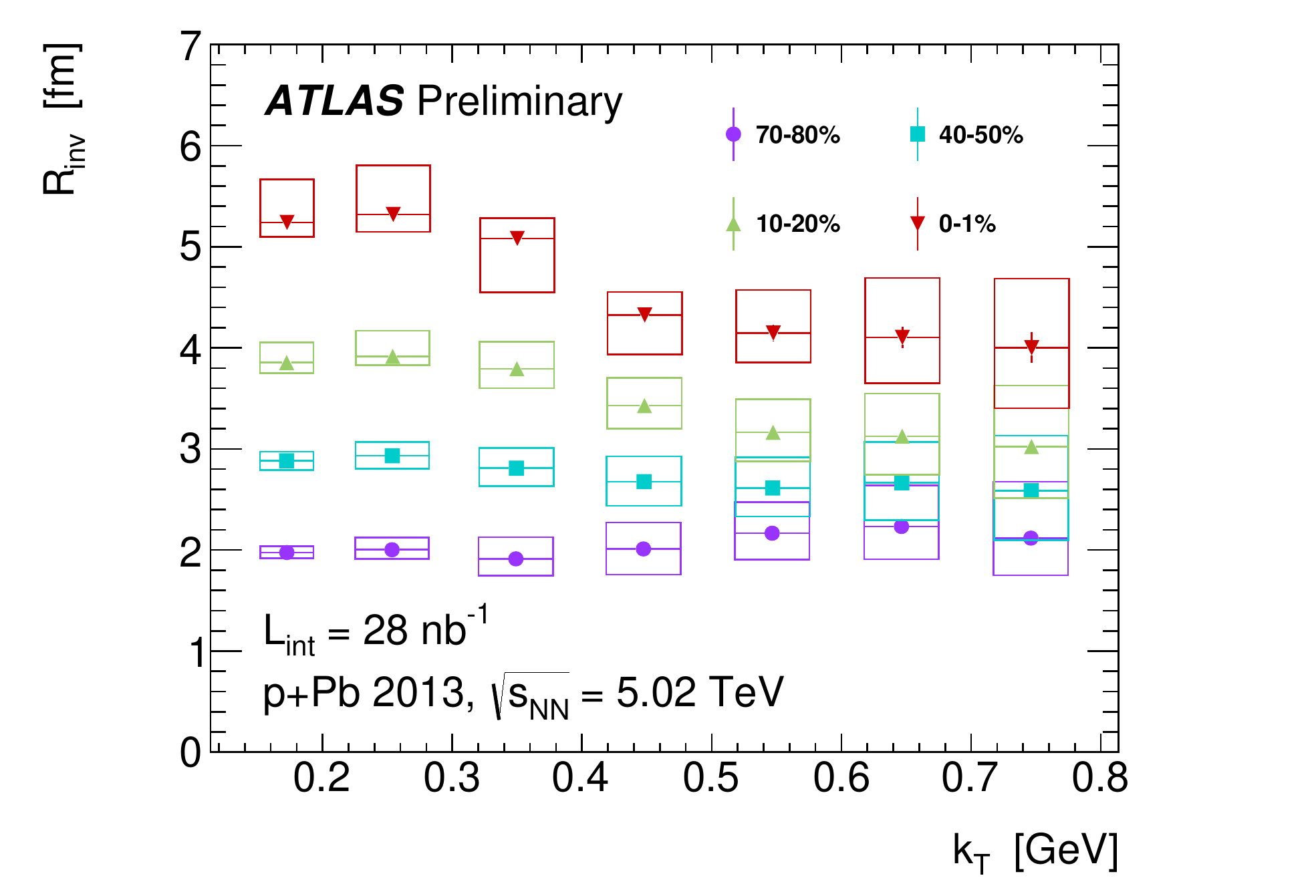}
\end{minipage}\hspace{3.5pc}%
\begin{minipage}{14pc}
\includegraphics[scale = 0.43]{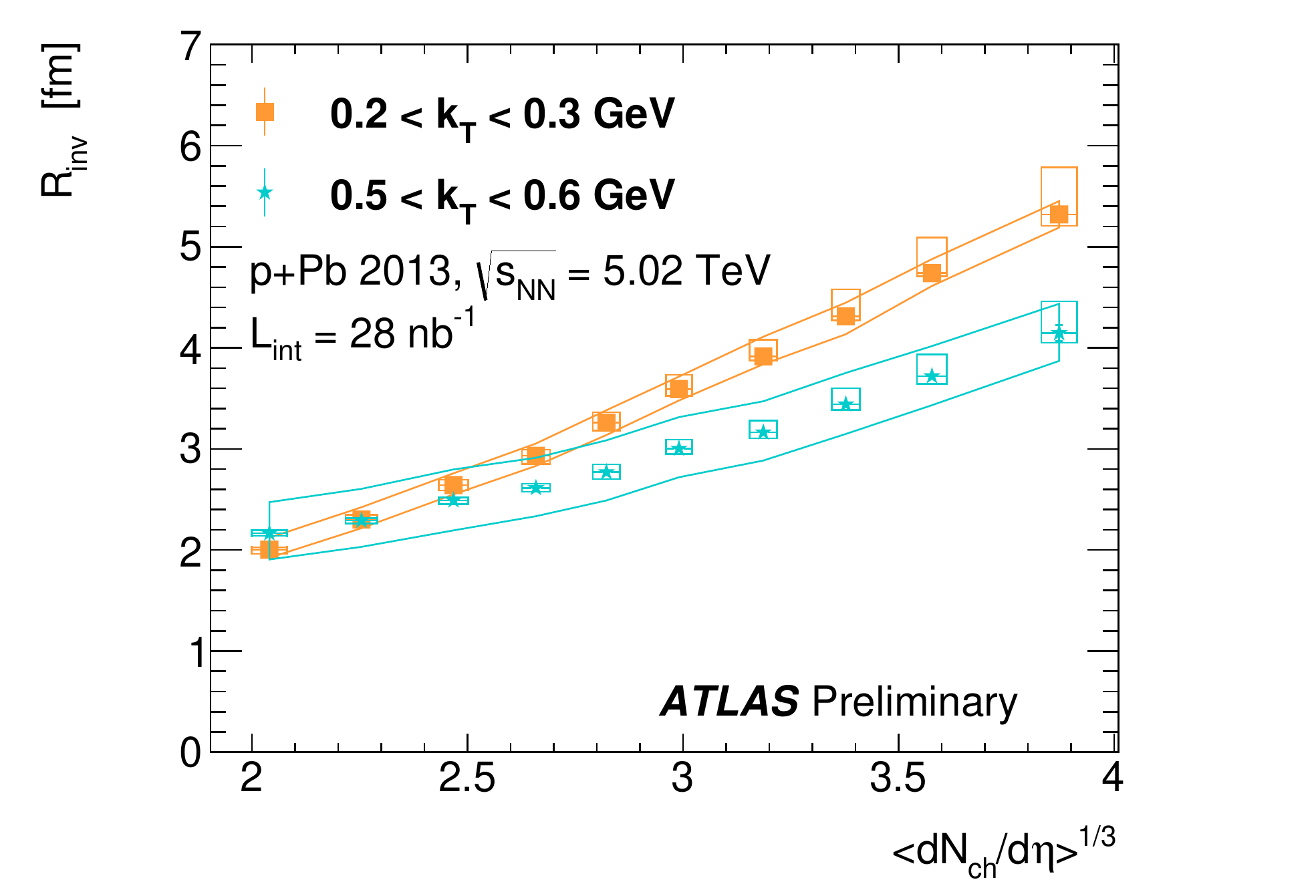}
\end{minipage}
\caption{Invariant radii $\rInv$ as a function of $\kT$ in different centrality bins (left panel) and as a 
function of the cube root of the average charged particle multiplicity, $\SqAvgMult$, in two $\kT$ intercepts (right 
panel). Figures from~\cite{ATLAS_femt_pPb_CONF}. }\label{fig:Rinv}
\end{figure}
\begin{equation}
 C (q) = \left[(1- \lambda) + \lambda K(q) C_{\text{BE} }(q) \right]  \Omega(q),
\end{equation}
where $\lambda$, $K(q)$, $C_{\text{BE} }(q)$ and $\Omega(q)$ is the correlation strength, a correction factor for 
final-state interactions, the Bose-Einstein enhancement factor and the contribution from non-femtoscopic correlations, 
respectively. The Bose-Einstein factor $C_{\text{BE} }(q)$ in the correlation function is fit to an exponential function
\begin{equation}
 C_{\text{BE} }(q) = 1 + \exp \left( - \hat{R} \hat{q} \right),
\end{equation}
where $\hat{R}$ and $\hat{q}$ are the invariant radius $\rInv$ and the invariant momentum $\qInv$ in 
the one-dimensional case and a diagonal matrix with the entries $\mathcal{R} = \text{diag}(\rOut,\rLong,\rSide)$ and 
$\vec{q} = (\qOut,\qLong,\qSide)$ in the three-dimensional case, respectively. \\
The non-femtoscopic contribution to the correlation function $\Omega(q)$ is estimated by a fit to the opposite-sign 
pair distribution
\begin{equation}
 \Omega(\qInv) = \mathcal{N} \left[ 1 + \lambda_{\text{bkg}} \exp \left(  - \left| R_{\text{bkg}} 
\qInv \right|^{\alpha_{\text{bkg}}} \right) \right], 
\end{equation}
where $\mathcal{N}$ is an arbitrary normalization factor, $\lambda_{\text{bkg}}$ is the experimental correlation 
strength, $R_{\text{bkg}}$ is a parameter for the width and $\alpha_{\text{bkg}}$ is a shape parameter. The origin of 
this background is identified to be from hard processes, such as particles from jet fragmentation. A mapping of the 
parameters $\lambda_{\text{bkg}}$ and $R_{\text{bkg}}$ between opposite-sign and same-sign is extracted from Monte 
Carlo simulations. It should be noted, that all parameters describing the background are constrained by this method.  
Systematic uncertainties are estimated by taking into account the hard process background description, particle 
identification, effective Coulomb correction size $R_{\text{eff}}$, charge asymmetry, and two particle effects. More 
details on the data analysis procedure can be found in Ref.~\cite{ATLAS_femt_pPb_CONF}.

\begin{figure}[!t]
\begin{minipage}{14pc}
\includegraphics[scale = 0.43]{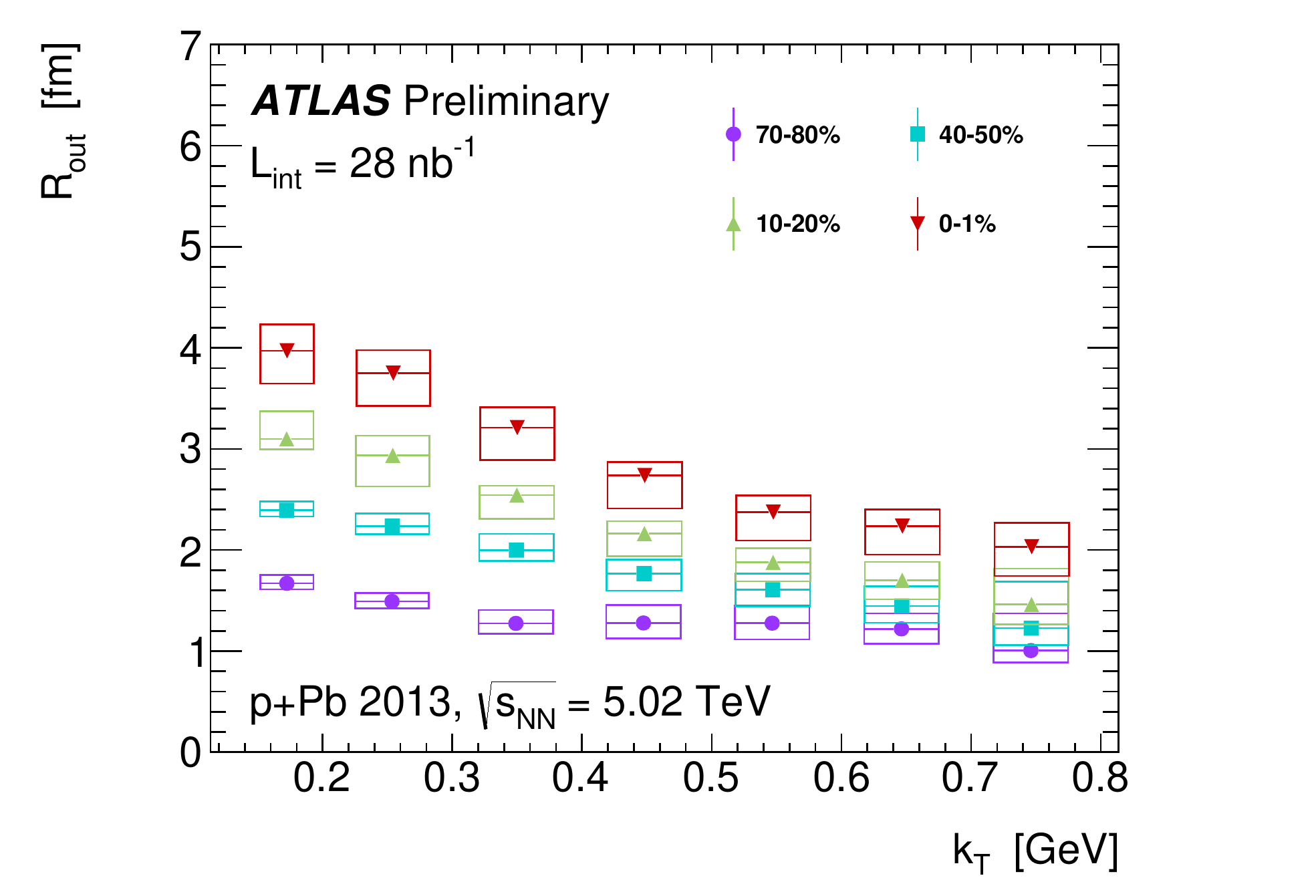}
\end{minipage}\hspace{3.5pc}%
\begin{minipage}{14pc}
\includegraphics[scale = 0.43]{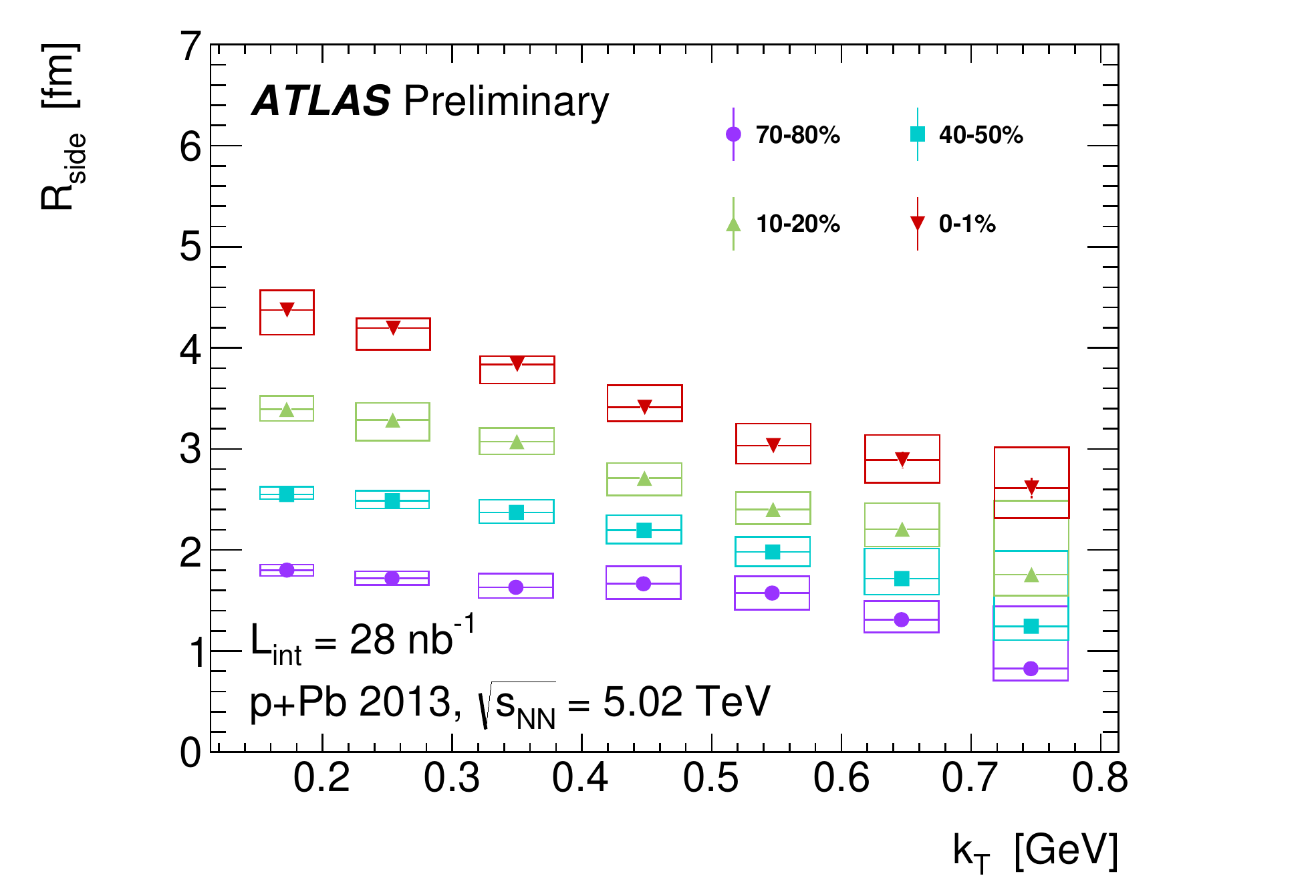}
\end{minipage}
\begin{minipage}{14pc}
\includegraphics[scale = 0.43]{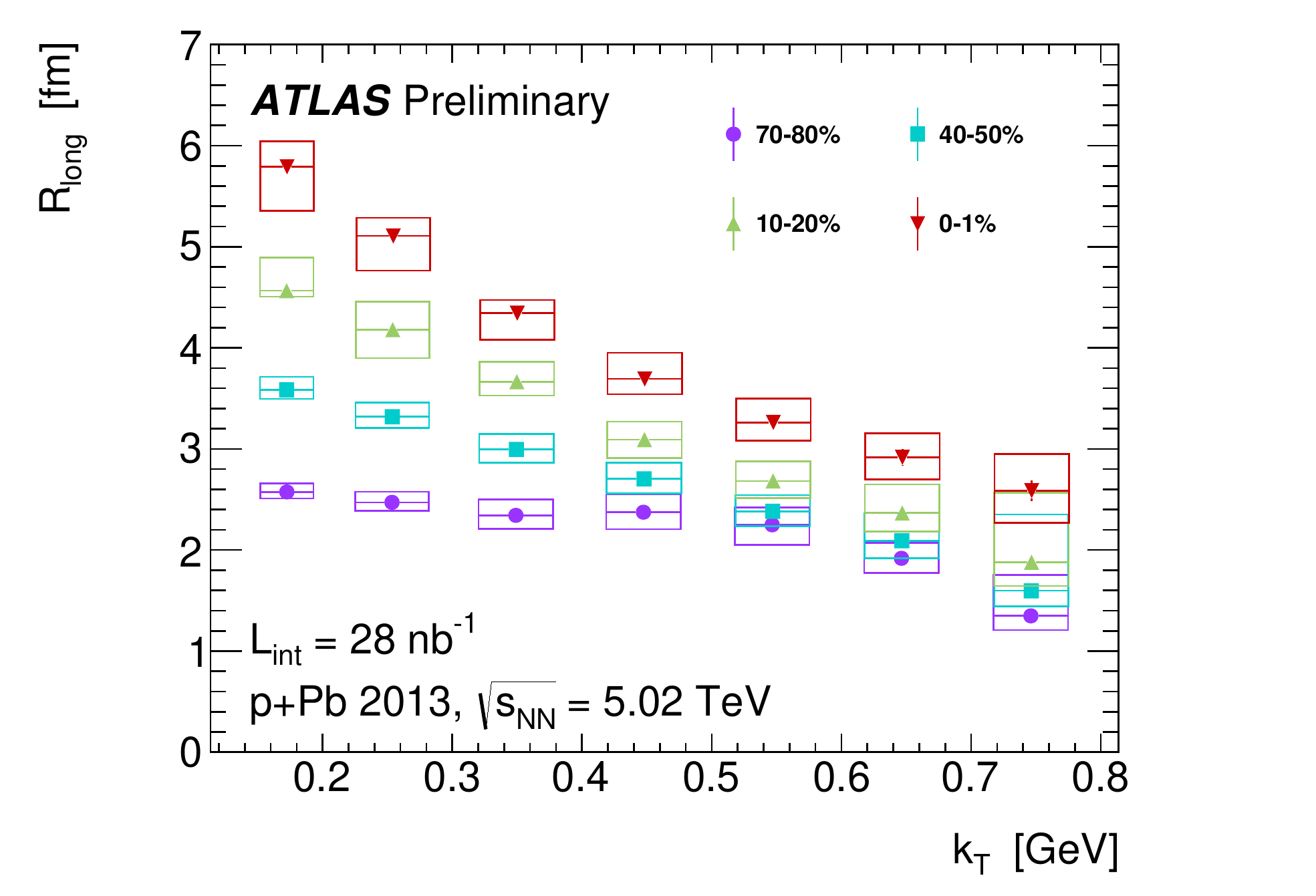}
\end{minipage}\hspace{3.5pc}%
\begin{minipage}{14pc}
\includegraphics[scale = 0.43]{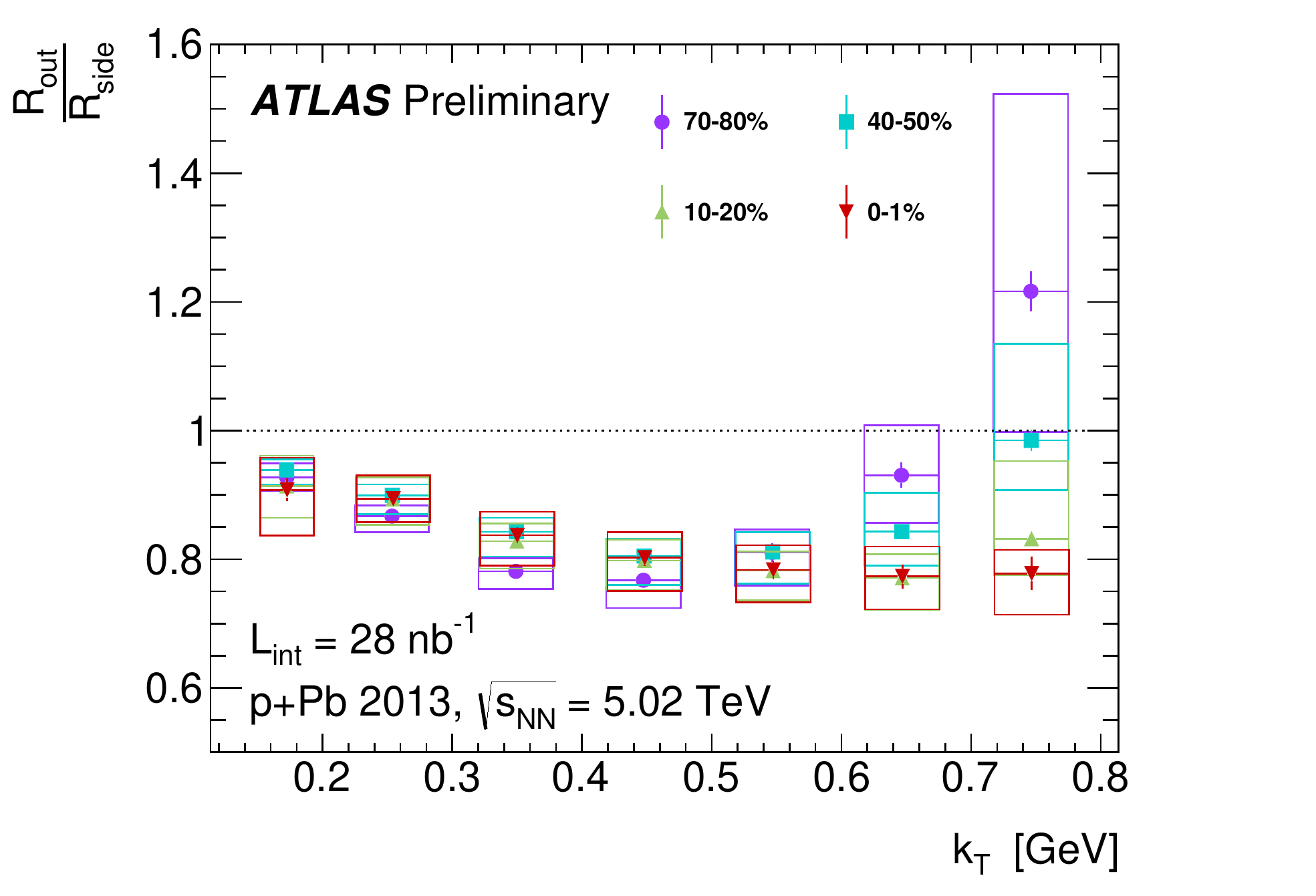}
\end{minipage}
\caption{The radii $\rOut$ (upper left panel), $\rSide$ (upper right panel) and $\rLong$ (lower left panel) as a 
function of $\kT$ in different centrality bins. The lower right panel shows the ratio of $\rOut/\rSide$ as function of 
$\kT$. Figures from~\cite{ATLAS_femt_pPb_CONF}. }\label{fig:3d}
\end{figure}
\section{Results and conclusions}
\label{sec:results_and_conclusions}
\begin{figure}[t]
\begin{minipage}{14pc}
\includegraphics[scale = 0.43]{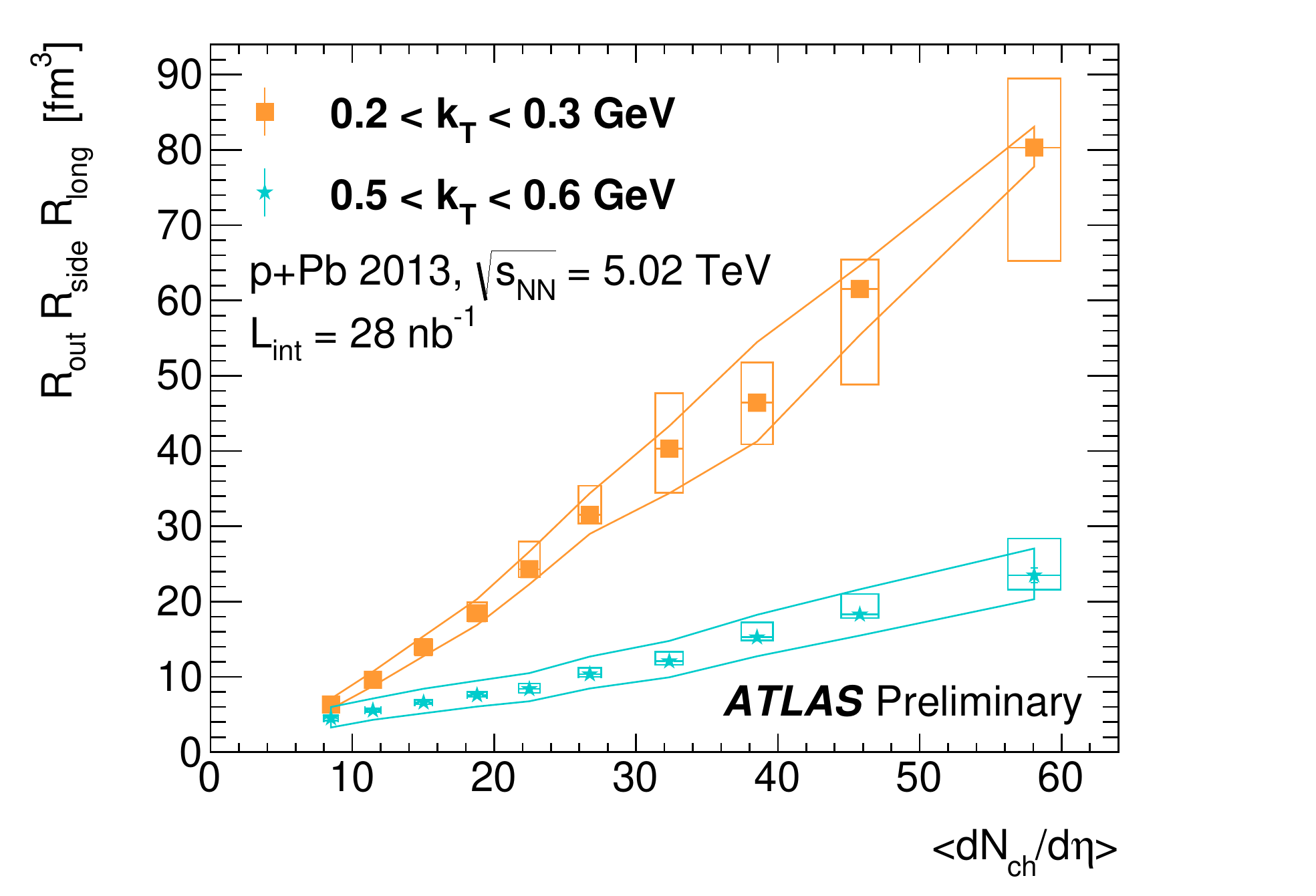}
\end{minipage}\hspace{3.5pc}%
\begin{minipage}{14pc}
\includegraphics[scale = 0.43]{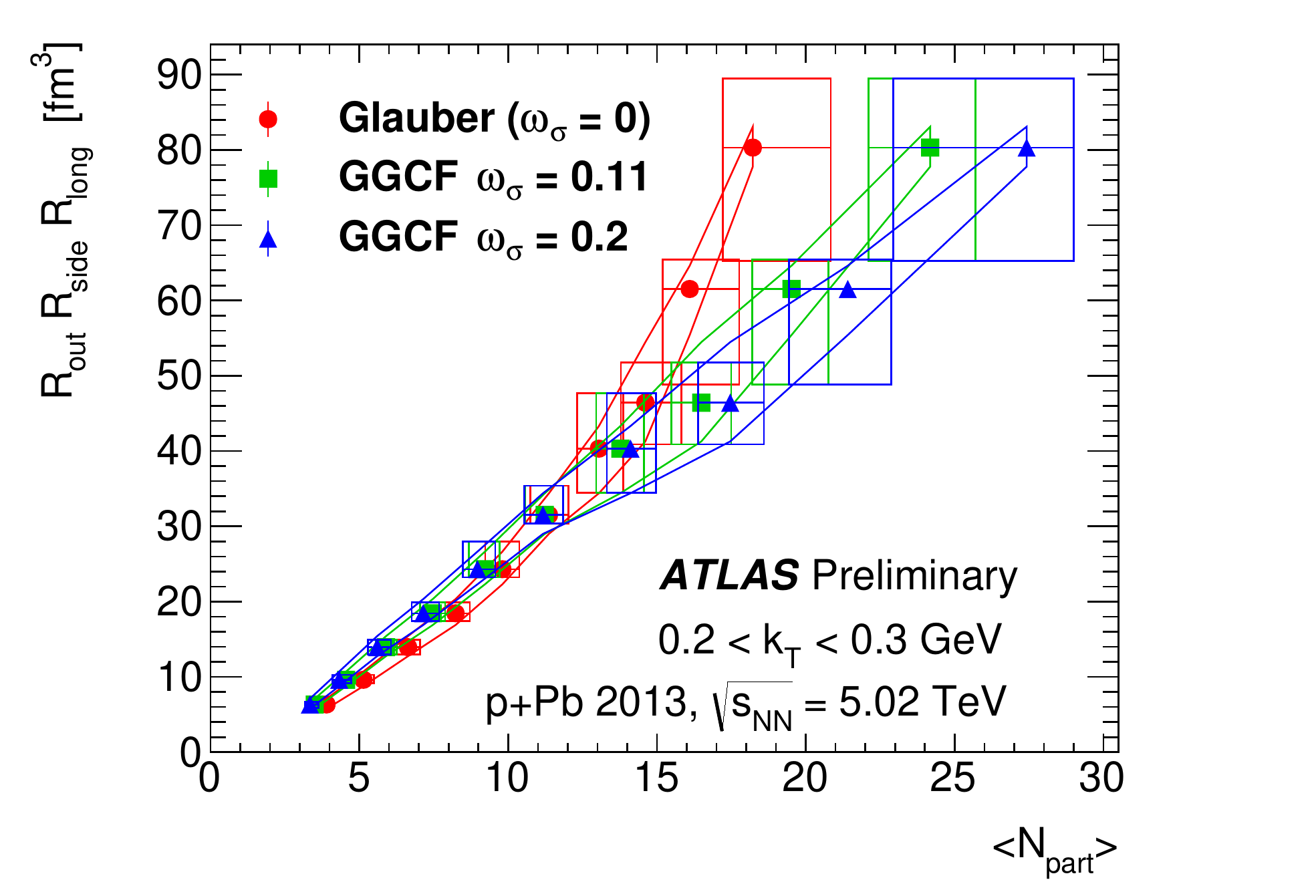}
\end{minipage}
\caption{The product $\rOut \rSide \rLong$ as a function of $\AvgMult$ for two different $\kT$ 
intervals (left panel) and as a function of $\avgNpart$ for three different models describing the 
initial geometry. Figures from~\cite{ATLAS_femt_pPb_CONF}.}\label{fig:HVolume}
\end{figure}

Figure~\ref{fig:Rinv} shows the invariant radii $\rInv$ as a function of the pair momentum $\kT$ and as a function of 
cube root of the average charged particle multiplicity $\SqAvgMult$. The measured radii are observed to decrease with 
increasing $\kT$, which is consistent with collective expansion. This behavior is less severe for peripheral 
collisions.\\
Figure~\ref{fig:3d} shows the results for the radii $\rOut$, $\rLong$ and $\rSide$ and the ratio 
$\rOut/\rSide$ as a function of the pair momentum $\kT$ for different centralities. The radii show a 
decreasing trend with increasing $\kT$. In the most central events $0\--1~\%$, the radii are about a factor $2.5$ 
larger than in peripheral collisions $70\--80~\%$. The ratio $\rOut/\rSide$ falls significantly below~$1$, indicating a 
very rapid 
and explosive expansion of the fireball. This behavior can be explained by a combination of pre-thermalized 
acceleration, a stiffer equation of state, and adding viscous corrections~\cite{Pratt09}.\\
In Fig.~\ref{fig:HVolume}, the product $\rOut \rSide \rLong$, which scales linearly with the volume, is 
shown as a function of the average multiplicity $\AvgMult$ for two intercepts of the pair momentum in the left panel. 
The volume is linearly increasing with increasing $\AvgMult$, indicating a constant source density at the moment of 
freeze-out. The product is also shown as a function of $\avgNpart$ for the Glauber Model and for the Glauber-Gribov 
Color Fluctuation model (GGCF), including different values $\omega_{\sigma}$ for the magnitude of the color 
fluctuations. The extraction of centrality dependent values of $\avgNpart$ is described in 
Ref.~\cite{ATLAS_cent_pPb}.\\ 
\section{Acknowledgement}
This work was supported by U.S. Department of Energy grant DE-FG02-86ER40281.








\bibliographystyle{elsarticle-num}






\end{document}